\documentclass[twocolumn]{aastex631}
\usepackage{mathtools} 
\usepackage{csvsimple}
\usepackage{longtable}
\usepackage{hyperref}

\usepackage{xcolor}
\newcommand*{\red}{\textcolor{black}}

\shorttitle{Are we blind to Helium exoplanets?}
\shortauthors{de Wit, Householder, \& Niraula}
\graphicspath{{./}{figures/}}
\usepackage{orcidlink}

\begin{document}

\title{Helium Atmospheres May Hide in Current Exoplanet Analysis Frameworks}

\author[0000-0003-2415-2191]{Julien de Wit}
\affiliation{Department of Earth, Atmospheric and Planetary Sciences, Massachusetts Institute of Technology, Cambridge, MA 02139, USA}

\author[0000-0002-5812-3236]{Aaron Householder}
\affiliation{Department of Earth, Atmospheric and Planetary Sciences, Massachusetts Institute of Technology, Cambridge, MA 02139, USA}
\affil{Kavli Institute for Astrophysics and Space Research, Massachusetts Institute of Technology, Cambridge, MA 02139, USA}

\author[0000-0002-8052-3893]{Prajwal Niraula}
\affiliation{Department of Earth, Atmospheric and Planetary Sciences, Massachusetts Institute of Technology, Cambridge, MA 02139, USA}
\affil{Kavli Institute for Astrophysics and Space Research, Massachusetts Institute of Technology, Cambridge, MA 02139, USA}


\begin{abstract}
The increasing number of detailed exoplanet observations offers an opportunity to refine our analyses and interpretations. Here, we show that \red{atmospheres that appear volatile-rich and/or cloudy} may instead be helium-rich. As transmission spectra constrain the atmospheric scale height ($H$), a He-enriched atmosphere can be misinterpreted as H$_2$-dominated water-rich to bring the mean molecular weight ($\mu$) to intermediate values ($\sim$4$-$10) when He/H$_2$ is fixed. \red{Similarly, a cloud deck can reduce the spectral features, and thus the apparent (i.e., cloud-free equivalent) $H$.}
We present a proof-of-concept reanalysis of  HD~209458~b's JWST transmission spectrum treating He/H$_2$ as a free parameter, resulting in sets of He-rich solutions. We argue that He enhancement must be considered to reliably constrain atmospheric composition, be sensitive to a more diverse planetary population, and ultimately yield robust trends to inform formation and evolution pathways. Looking ahead, we suggest leveraging insights from differences in pressure-broadening effects, outflow measurements, and atmospheric chemistry to disentangle reliably between He-, volatile-rich\red{, and cloudy} atmospheres---while recognizing that \red{associated} models need targeted upgrades to reach the fidelity level \red{required to this end}. 

\end{abstract}


\section{Introduction}
The field of exoplanetary science has undergone a remarkable transformation in the last decades, evolving from the detection of a handful of exoplanets to the \red{increasingly}-detailed characterization of exoatmospheres. Yielding data with unprecedented information content, JWST is at the forefront of this transformation. 

As the community learns to leverage and interpret the novel data, fresh challenges arise. Reliably translating a planet's transmission spectrum into physical constraints now requires exquisite knowledge of its host's heterogeneities \citep[e.g.,][]{Lim2023,Moran2023,Rackham2023,Berardo2024,TJCI2024}, and of light-matter interactions \citep[e.g.,][]{Fortney2019, Greaves2021, Ranjan2020, Niraula2022}. Beyond model upgrades, it is also key to keep (re)assessing the complexity we allow for in our models to ensure they reach a sufficient fidelity. For example, using simple 1D models can prevent the identification of day-night temperature asymmetries and lead to underestimated temperatures by up to $\sim$1,000K \citep{MacDonald2020}. 

Here, we examine the impact of a common assumption behind the interpretation of giant exoplanets' spectra: a helium-to-hydrogen (He/H$_2$) ratio identical to that of Jupiter \citep[He/H$_2$ = 0.157,][]{Conrath1987}. As most JWST datasets target giant exoplanets under irradiations beyond those found in the solar system, freeing atmospheric retrieval codes from an assumption anchored in solar-system observations regarding He/H$_2$ seems well motivated. \citet{Hu2015} first presented the hypothesis of a He-dominated atmosphere for a Neptune-sized planet (GJ~436~b possibly turned He-rich via atmospheric hydrodynamic escape). Since other studies have investigated a range of fractionation regimes for similar and smaller planets \citep[e.g.,][]{Malsky2023,Cherubim2024,Lammer2025}.

We present our concept in Section~\ref{sec:concept} and a proof-of-concept using HD~209458~b's JWST spectrum in Section~\ref{sec:data}. We explore possible origins for and means to disentangle between He-rich, \red{volatile-rich, and cloudy} atmospheres in Section~\ref{sec:discussion} and conclude in Section~\ref{sec:conclusion}. 

\section{Concept}
\label{sec:concept}

A planet's transmission spectrum can yield independent constraints on its temperature, composition, and pressure profile \citep{deWit2013}, including its atmospheric scale height \citep[$H = k T/\mu g$, with $k$ Boltzmann's constant, $T$ the temperature, $\mu$ the molecular weight, and $g$ the gravity---see also][]{Miller2009}. This can result in tight constraints on the atmospheric mean molecular weight \citep[e.g.,][]{Niraula2025,dewit2025}, notably for planets with independent mass constraints from radial velocities or transit timing variations. Therefore, if $\mu$ deviates from the default $\sim$2.3 due to helium enrichment (or the presence of a dominant atmospheric species without strong absorption features--i.e., not accounted/fitted for in a retrieval), retrieval codes artificially amplify the abundance of the dominant absorber (often water)\red{---or involve clouds \citep[see Section~\ref{sec:others} and, e.g.,][]{Berta2012}}. 
 
\begin{figure}[t!]
    \centering
    \includegraphics[width=0.50\textwidth, trim={1cm 1cm 0cm 1cm},clip,]{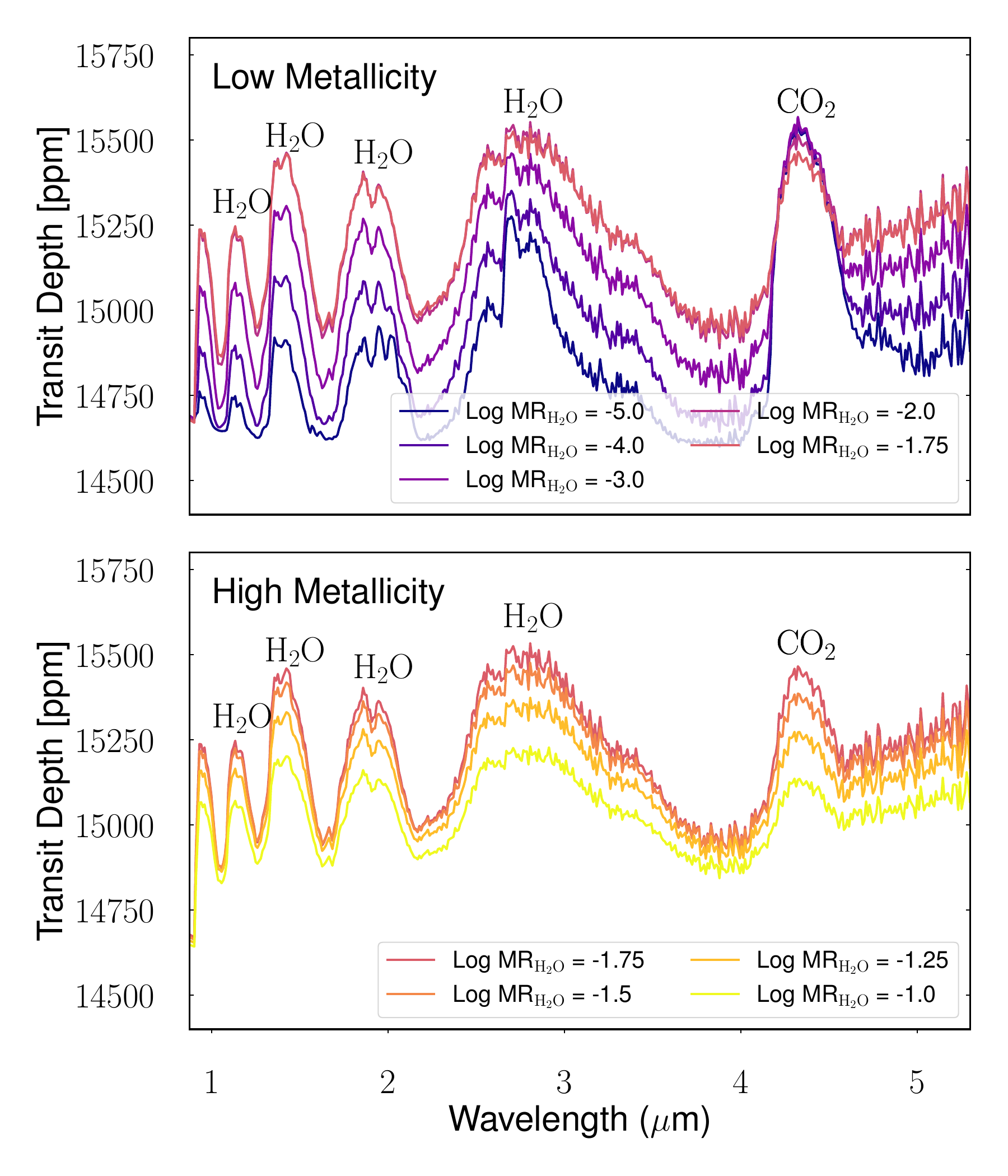}
    \caption{The two regimes of effects of water abundance on a transmission spectrum. At low water abundances (MR$_{\rm H_2O} \lesssim 1\%$), increasing the amount of water primarily increases $\rm H_2O$ absorption features while leaving the rest of the spectrum mostly unchanged. At high water abundances (MR$_{\rm H_2O} \gtrsim 5\%$), increasing the water content primarily raises the atmospheric mean molecular weight, shrinking the atmospheric scale height and compressing the entire spectrum. This effect is expected for other volatiles and can be generalized to ``volatile-rich'' atmospheres.}
    \label{figure:differentmetallicity}
\end{figure}

To illustrate this, we show the two main regimes of effects an absorber's abundance have on a transmission spectrum in \autoref{figure:differentmetallicity}: (1) at ``low'' abundance the absorption features of the absorber increase with its abundance, (2) at ``high'' abundance all absorption features decrease with abundance. \autoref{figure:differentmetallicity} presents a set of synthetic transmission spectra for a giant exoplanet for which only the atmospheric water vapor abundance is changed (log volume mixing ratios (VMRs) of H$_2$O ranging from $-5.0$ to $-1.0$). We adopt properties consistent with HD~209458~b (incl., mass, radius, equilibrium temperature), and assume an isothermal, cloud-free atmosphere. At low abundances (i.e., $\lesssim 1\%$ H$_2$O), the abundance mainly affects the strength of H$_2$O absorption features. At higher abundances (i.e., $\gtrsim 1\%$ H$_2$O), the H$_2$O features are saturated, and the H$_2$O abundance predominantly influences the overall amplitude of the transmission spectrum by increasing $\mu$ and thus decreasing $H$.

We thus hypothesize that high water content (or, in fact, metallicity) may result from biases stemming from assuming He/H$_2$ = 0.157. This forces the dominant absorbers with saturated absorption features to higher abundances to tweak $\mu$ and match the $H$ value recorded by the transmission spectrum. This would mean that current (fixed-He/H$_2$) retrieval frameworks are blind to He-rich atmospheres and risk yielding artificially enhanced metallicity trends. \red{We discuss in Section~\ref{sec:others} that the same goes for clouds.}

\vspace{-0mm}
\vspace{-0mm}

\begin{figure*}
    \hspace{-10mm}
    \includegraphics[width=1.07\textwidth]{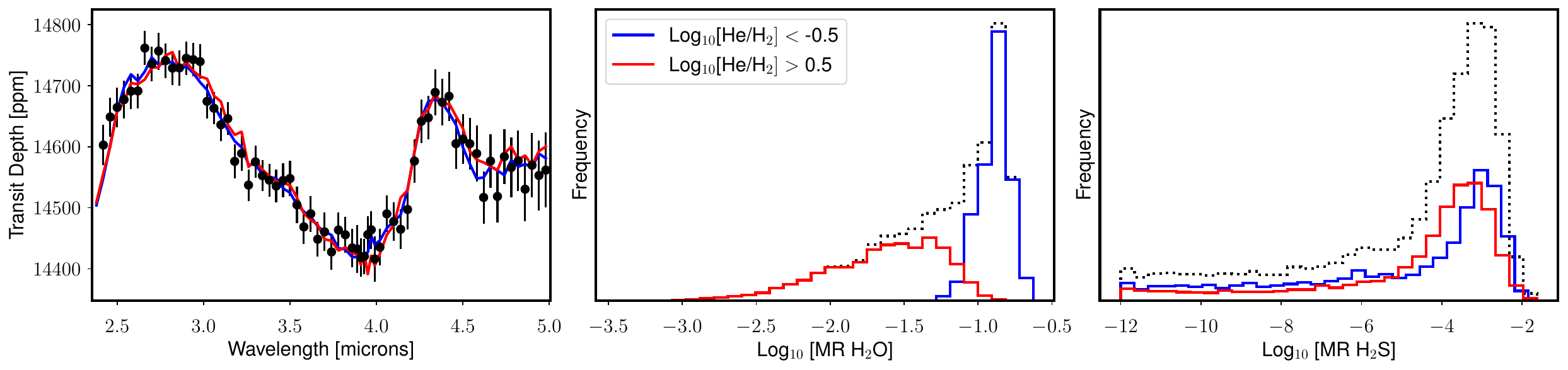}
    \caption{JWST's spectrum of HD~209458~b is consistent with water-rich or He-rich \red{cloud-free} atmosphere. \textbf{Left:} Best fits to HD~209458~b's transmission spectrum from the Log$_{10}$[He/H$_2$]$<-0.5$ (blue) and $>0.5$ (red) subsample of solutions. Data from \citet{Xue2024}. \textbf{Center:}  Posterior probability distributions (PPDs) for water abundance under low (red) and high (blue) He/H$_2$ ratios. Higher helium enrichment leads to lower inferred H$_2$O, since the increased mean molecular weight alone can match the scale height of the atmosphere. \textbf{Right:} The same comparison for the trace gas H$_2$S shows negligible differences between the two subsamples, as non-dominant absorbers are less sensitive to assumptions about bulk atmospheric composition.}
    \label{figure:bestfit}%
\end{figure*}

\section{Proof of Concept}
\label{sec:data}

\subsection{Data}
To test this idea, we reanalyze HD~209458~b's JWST spectrum while fitting for He/H$_2$.
The spectrum used is derived from two 8.01-hour transit observations with JWST-NIRCam (GTO 1274; PI: Lunine) and was reduced using \texttt{Eureka} \citep{Bell2022}. We note that each of the reported uncertainties were inflated by a factor of $\sim$1.4 to account for underestimated noise, which explains the apparent lack of large residuals in the model fits to the data (see \autoref{figure:bestfit}).

\subsection{Analysis Framework}

\label{3}

To perform our atmospheric retrievals, we use \texttt{tierra}\footnote{\href{https://github.com/disruptiveplanets/tierra}{https://github.com/disruptiveplanets/tierra}} \citep{Niraula2022,Niraula2023,Niraula2025}. Unlike many previous retrievals, we treat He/H$_2$ as a free parameter to assess its impact on the retrieved abundances, and test for He enrichment or depletion. \red{For this first proof-of-concept application, we assume the atmosphere to be cloud-free.}

For our retrievals, we consider eight different molecules: $\mathrm{CH_4}$, $\mathrm{CO}$, $\mathrm{CO_2}$, $\mathrm{HCN}$, $\mathrm{H_2O}$, $\mathrm{H_2S}$, $\mathrm{NH_3}$, and $\mathrm{SO_2}$, applying a uniform prior on the log mixing ratio ranging from -0.5 to -12, and check later that none of the PPDs are affected by this choice (i.e., the retrieved molecular abundances did not pile up at the upper bounds of the prior, \red{see Section~\ref{sec:others} }).  We also vary the planetary mass, placing Gaussian priors based on the published radial velocity mass \citep{Bonomo2017}.

\subsection{Results}
\label{sec:results}

Our \red{cloud-free} retrievals return a broad range of possible values for He/H$_2$ and a clear negative trend between He/H$_2$ and the H$_2$O abundance (\autoref{figure:bestfit}).  
As anticipated (Section\,\ref{sec:concept}), the increasing He/H$_2$ ratio also lessens the reliance on H$_2$O abundance in increasing $\mu$, which  results in H$_2$O's abundance being less constrained then. 

\subsubsection{A Perspective from Molecular Abundances}

To highlight this further, we split our ensemble of solutions into those with Log$_{10}$[He/H$_2$]$<-0.5$ and $>0.5$  in \autoref{figure:bestfit}. This helps show that retrievals with higher He/H$_2$ ratios consistently yield lower water abundances, while lower He/H$_2$ ratios constrain water abundance to higher values and a much tighter part of the parameter space (\autoref{figure:bestfit}, middle panel).  Specifically, we find a water abundance of $9.1^{+2.1}_{-1.7}\%$ for the Log$_{10}$[He/H$_2$]$<-0.5$ sample and $2.95^{+2.2}_{-2.4}\%$ for Log$_{10}$[He/H$_2$]$ > 0.5$. 

Conversely, the abundance of a non-dominant absorber like H$_2$S is affected marginally by higher He/H$_2$ ratios.  H$_2$S contributes negligibly to $\mu$ compared to more abundant species with saturated absorption features like H$_2$O. As a result, when the He/H$_2$ ratio increases and $\mu$ rises, the retrieval framework does not need to significantly adjust/compensate the inferred H$_2$S abundance to maintain a good fit.

\begin{figure}
  \centering
  \includegraphics[width=0.45\textwidth]{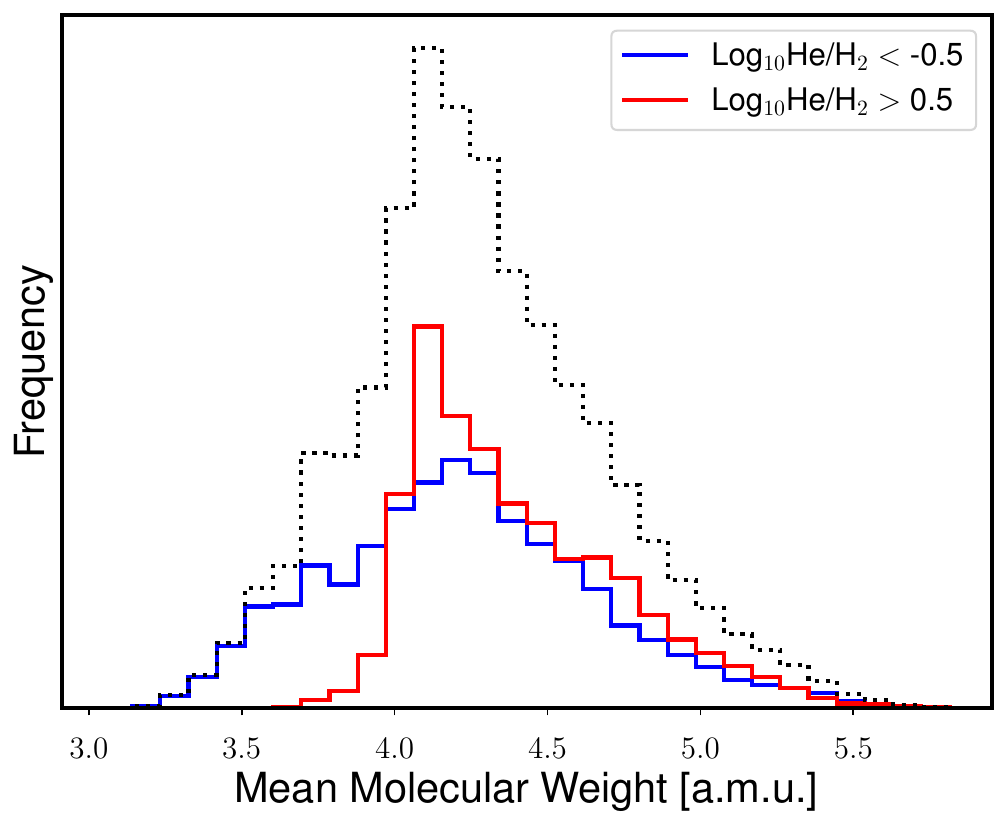}
  \caption{Normalized posterior distribution of the mean molecular weight $\mu$ for \red{cloud-free} HD\,209458~b peaking above the canonical $\mu \simeq 2.3$ expected for a He/H$_2$ ratio of 0.157.}
  \label{fig:mu}
\end{figure}

\subsubsection{A Perspective from the Mean Molecular Weight}

We can also look at this matter through the lens of $\mu$. Transmission spectra primarily constrain $\mu$ through their sensitivity to $H$ \citep{Miller2009,deWit2013}. In our retrievals, we independently computed $\mu$ by summing the retrieved mixing ratios of all species, weighted by their molecular masses. The peak of the distribution lies significantly beyond the canonical value of $\mu \simeq 2.3$ expected for a Jupiter-like composition (\autoref{fig:mu}), suggesting a substantially-heavier atmosphere. If one instead assumes a fixed $\mu=2.3$, the retrieval must compensate by artificially inflating the abundance of the dominant absorber (typically water). This degeneracy highlights how fixed assumptions about He/H$_2$ can directly lead to overestimates of atmospheric water content (or metallicity).

\section{Discussion}
\label{sec:discussion}

\subsection{Giant exoplanets with Helium atmospheres?}

While He-rich giants may sound exotic at first, He-rich atmospheres require $\lesssim$1 order of magnitude enrichment \red{ compared to the gas giants of the solar system. This is over an order of magnitude less enrichment than needed for a volatile-rich giants (e.g., $O(1\%)$ water)}. Although hydrogen is the most abundant element in solar-system gas giants, it is conceivable that a giant exoplanet becomes He-rich---e.g., under conditions where hydrogen is more readily photoevaporated. 

\begin{figure}[ht!]
    \hspace{-10mm}\includegraphics[width=0.58\textwidth]{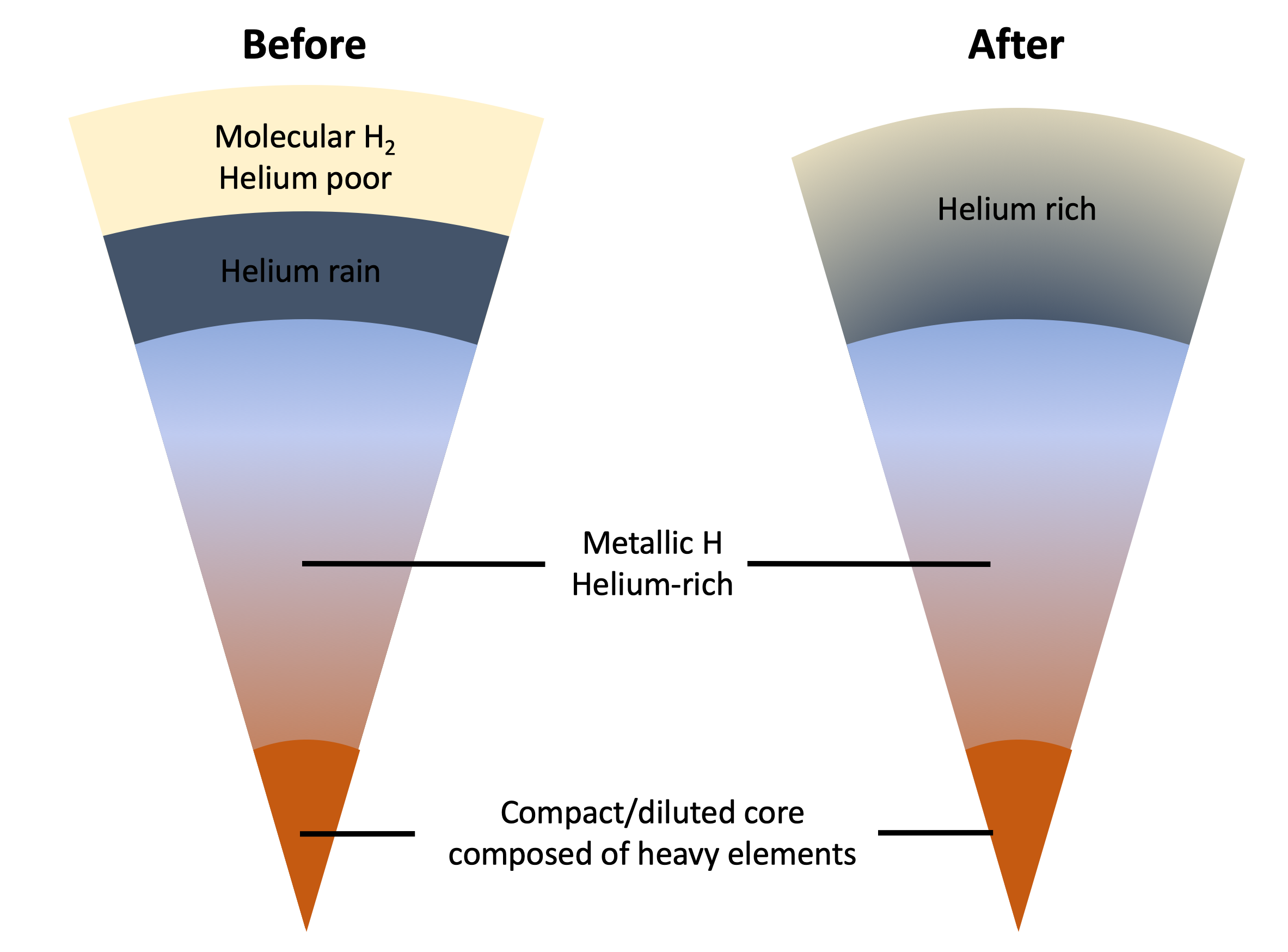}
    \caption{Forming a He-dominated atmosphere for a giant exoplanet. Schematic representation of the ``before'' and ``after'' states of atmospheric evolution for a stratified giant atmosphere -- e.g., via helium rain. He fractionation is enhanced via the formation of a H$_2$-dominated He-poor layer at the top of the atmosphere, allowing for the preferential loss of H$_2$ and preservation of He as seen for HD~209458~b, for example \citep{Xing2023}.}
    \label{figure:evolution}
\end{figure}

Figure~\ref{figure:evolution} illustrates a schematic ``before and after'' evolution for a giant atmosphere having undergone such an enrichment. In the \red{pre-migration} initial state (left), the planet has a stratified structure with a molecular hydrogen-dominated, helium-poor upper atmosphere, a helium rain layer where helium droplets condense and sink. Over time, as hydrogen is selectively lost from the upper layers via photoevaporation \citep{Owen2013,Owen2017}, the planet transitions to the state shown on the right: a helium-rich atmosphere where the helium-enriched regions dominate the transmission spectroscopy signal with an intermediate $\mu$ and thus a lower $H$. This mechanism may explain why some exoplanets, especially Neptune-sized and smaller \citep[e.g.,][]{Hu2015,Malsky2023,Cherubim2024,Lammer2025}, might exhibit atmospheres more He-rich than Jupiter's or Saturn's.

\red{While an atmosphere-only helium enrichment (or inverted metallicity gradients) may at first be predicted as unstable \citep[e.g.,][]{Thorngren2019}, new mechanisms have been introduced to explain the newly-discovered inverted metallicity gradient in Jupiter \citep{Muller2024}---e.g., a stronger subadiabaticity to stabilize the outer stable layer and support, notably an inverted helium gradient \citep{Nettelmann2025}.} The alternative of a (quasi-)global helium enrichment of the planet presents the same challenge at first: interior models predicts Jupiter-mass objects composed primarily of helium to be as small as $\sim0.5R_{Jup}$ \citep[e.g.,][]{Stevenson1976}. However, the same models predicts radii up to 2x smaller than observed for a subpopulation of hot Jupiters \citep[a phenomenon coined ``the mystery of the inflated radii'', see][and references therein]{Thorngren2018}. In other words; to avoid biases and ensure that we fully harvest the potential of this data-rich era of astronomy, it is important to appreciate that---while helpful---models are still subject to observation-driven upgrades, as recently highlighted by the discovery of a helium-rich (and carbon-rich) ``planet'' \citep{Zhang2025}.

\vspace{-0mm}
\vspace{-0mm}

\vspace{-0mm}
\begin{figure*}[ht!]
    \hspace{-2mm}\includegraphics[width=0.98\textwidth, trim={0cm 0cm 0cm 0cm},clip,]{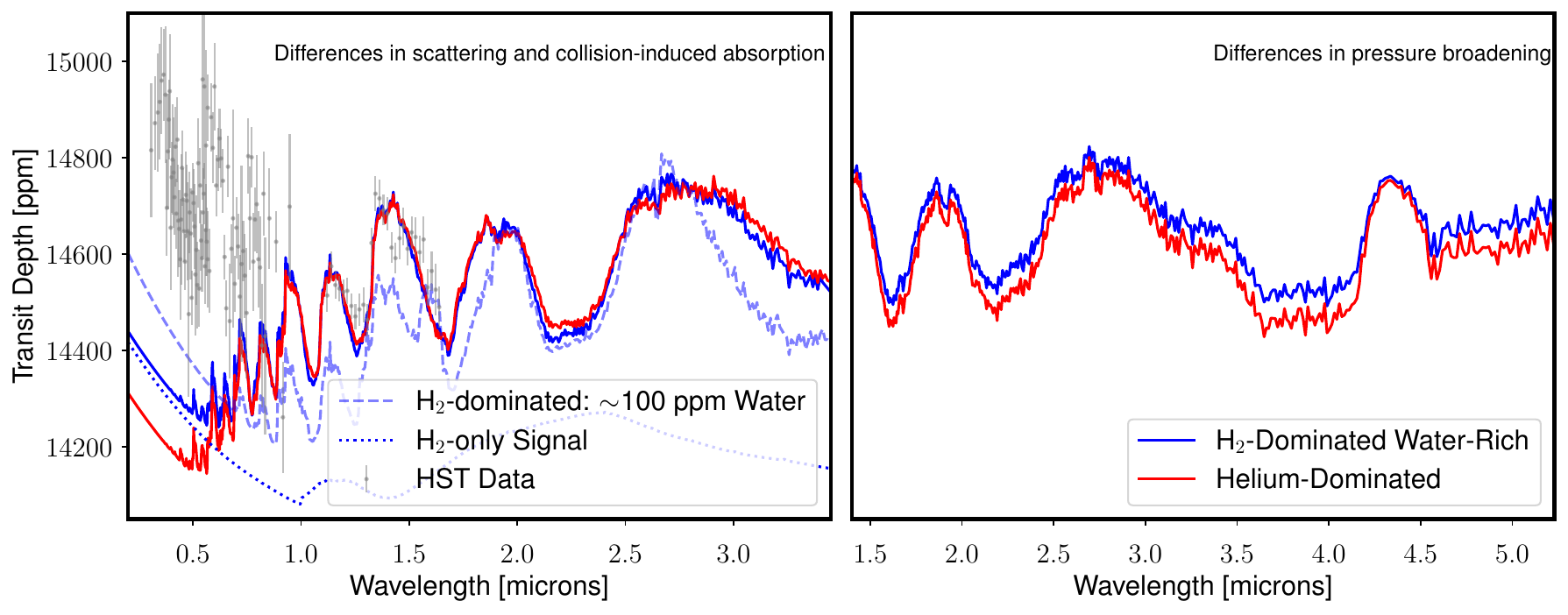}
    \caption{Spectroscopic differences between a H$_2$- and He-dominated atmosphere. Left: Best fits to HD~209458~b's transmission spectrum from Fig.\ref{figure:bestfit} (Log$_{10}$[He/H$_2$]$<-0.5$ in blue and $>0.5$ in red) propagated on a wider wavelength range to present where the H$_2$- and He-dominated scenarios diverge, esp. due to scattering ($\lesssim1\mu$m) and collision-induced absorption (CIA, $\sim2.4\mu$m) -- see dotted blue line for the H$_2$-only signal. While differences in scattering can yield $\sim$100-ppm signals, hazes will often overpower the scattering of H$_2$/He \citep[HST data for HD~209458~b in grey, ][]{Sing2016}. Similarly, for volatile-rich atmospheres, H$_2$'s CIA signal will not shine between features of, e.g., H$_2$O. (In comparison, H$_2$'s CIA signal is clearly detectable for $\lesssim$100\,ppm water abundance, dashed blue line.)  Right: Synthetic transmission spectra from the best-estimate parameters for HD~209458~b using perturbed cross-sections with doubled and halves pressure-broadening coefficients from \citet{Niraula2022} to simulate the detectable effects of H$_2$- vs He-broadened absorption features of other gases.}
    \label{fig:rayleighscattering}
\end{figure*}
\vspace{-0mm}

\begin{figure}
  \centering
  \includegraphics[width=0.45\textwidth]{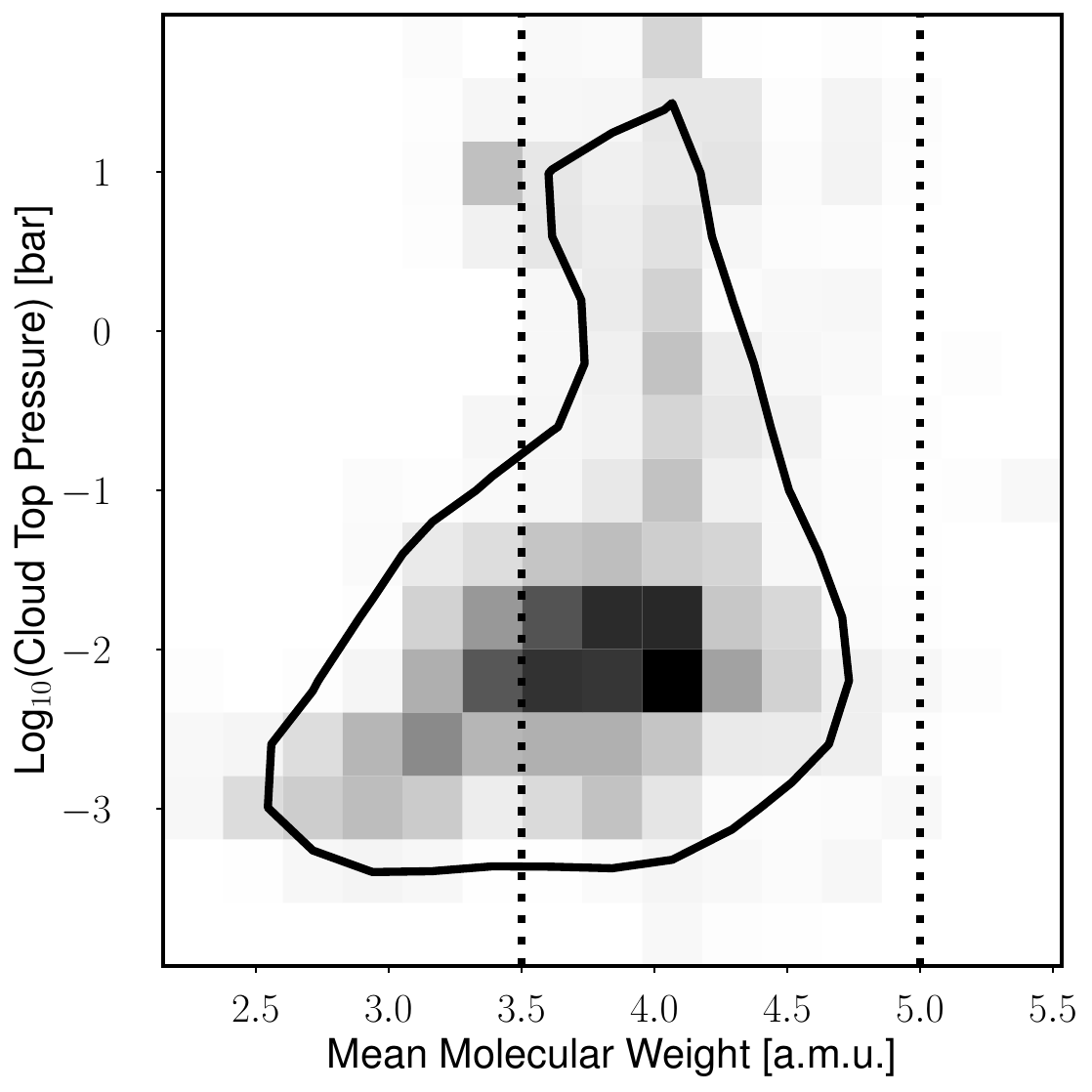}
  \caption{\red{Posterior probability as a function of $\mu$ and cloud-top pressure for HD~209458~b (95\% contour shown). The dotted lines present the 95\%-confidence interval from the cloud-free analysis (see Fig.~\ref{fig:mu}). For solutions with cloud top above $\sim$100~mbar, He/H$_2$ ratios increasingly consistent with the canonical $\mu \simeq 2.3$ are allowed.}}
  \label{fig:mu_cloud}
\end{figure}

\subsection{Other Possible Explanations}
\label{sec:others}

While we present in this work the possibility of an alternative interpretation to the some giant exoplanets' recent JWST spectra, it is important to recognize that an ensemble of other possible explanations exists. 

First and foremost, these worlds may very well be truly water-rich (see set of blue solutions in Figure\,\ref{figure:bestfit}). Although such high abundances are not anticipated based on solar system giants or equilibrium chemistry models \citep[e.g.,][]{Molliere2017}, missing physics and chemistry in current formation or atmospheric evolution models could allow for heavier-than-expected compositions in highly irradiated environments.

Second, another spectrally inactive background (not He) with a high mean molecular weight could also result in the same effect (while requiring a lower abundance). Like helium, these spectrally inactive species could raise $\mu$ and $H$, prompting retrieval frameworks to compensate the models by inflating the abundance of a strong absorber with saturated absorption features such as water. We consider the latter alternative less likely considering the relative abundance of He vs C, N, O, etc. which would lead to volatile-rich atmospheres before N$_2$- or Ne-rich atmospheres, for example.

\red{Third, reduced spectral features can also result from the presence of high-altitude clouds \citep{Berta2012}. To showcase this additional dimension of relevance, we relax the could-free assumption in our analysis of HD~209458~b's spectrum. Doing so, we find that for cloudy scenarios (i.e., with cloud top above $\sim$100~mbar) $\mu$ is not constrained to intermediate values, but can be consistent with conventional He/H$_2$ ratios (Fig.~\ref{fig:mu_cloud}).}

\subsection{He-rich? Volatile-rich? Which is which?}
\label{sec:future}
While JWST offers unparalleled precision for constraining atmospheric compositions, directly constraining the He/H$_2$ ratio for giant exoplanets is challenging. We discuss here three sources of complementary information to help disentangle between \red{the He- vs volatile-rich scenarios---see \citet{Benneke2013} for further details on unambiguously identifying cloud signatures}. 

\subsubsection{Scattering, Absorbing, and Broadening Behaviors}

The first avenue to directly identify the dominant constituent in an atmosphere is to identify absorption/attenuation features. For the present H$_2$-vs-He scenario, we can rely on (1) the Rayleigh slope, (2) Collision Induced Absorption (CIA) for H$_2$, and (3) their respective pressure-broadening effect. 

To highlight these three spectral effects and showcase how they may help break the degeneracy between He-rich and (H$_2$-dominated) water-rich atmospheres, we present in the left panel of \autoref{fig:rayleighscattering} the extended best-fit models to the HD~209458~b spectrum presented in Figure\,\ref{figure:bestfit}. Rayleigh scattering values are taken from \citet{wilmouth2019} and \citet{sneep2005} for helium and hydrogen, respectively. It shows that the best fit models deviate substantially from each other in the VIS/NIR, primarily due to He's scattering being $\sim$2 orders of magnitude smaller than H$_2$'s. In clear atmospheres, it would thus be possible to identify the differences between two scattering signatures. Unfortunately, close-in planets often have cloudy/hazy atmospheres supported by enhanced photochemical production \citep[see HST data for HD~209458~b in \autoref{fig:rayleighscattering} and, e.g.,][]{Sing2016}. We thus expect that disentangling between a helium- and hydrogen-dominated atmosphere solely from the scattering signal to be challenging.

The next clear difference between both gases relates to the presence/absence of H$_2$ CIA features. We point in particular to the $\sim$2.4$\mu$m feature able to peak between water and methane features (\autoref{fig:rayleighscattering}, left panel). Relative measurements of out-of-feature transit depths (i.e., 2.4$\mu$m vs 1.7 and 3.4 $\mu$m) can provide direct insights into the contribution of H$_2$'s CIA and leverage part of transmission spectra less sensitive to the signals that can overpower scattering slopes \citep[e.g., hazes and stellar contamination, see][]{Triaud2024}. Unfortunately, the aforementioned out-of-feature windows will close for volatile-rich atmospheres (\autoref{fig:rayleighscattering}, left panel), preventing the detection H$_2$'s CIA.

Finally, we turn to the secondary spectroscopic effect associated the pressure-broadening of other species' lines. \citet{Niraula2022} demonstrated that information (or the lack thereof) associated with pressure broadening in central to an exoplanet transmission spectrum, while \citet{Wiesenfeld2025} reported that JWST-quality spectra can be sensitive to difference in the pressure-broadening coefficients down to $\sim$10\%. Fortunately, the difference between the pressure-broadening coefficients of most gases by H$_2$ vs by He is one order of magnitude above that sensitivity limit \citep[e.g., ][and references therein]{Tan2022}. We showcase in the right panel of \autoref{fig:rayleighscattering} the detectable differences in the transmission spectrum of an exoplanet for which only the pressure-broadening coefficients have been changed; $\sim$50-ppm level across the spectrum, esp. in the out-of-feature regions (more pressure-broadening sensitive). As retrievals can hide such a signal, reliably sensing and interpreting such H$_2$-vs-He spectral differences will require targeted upgrades of opacity models \citep{Tan2022,Niraula2022,Wiesenfeld2025}.

\subsubsection{Outflow Fractionation}

Another promising avenue to help disentangle the H$_2$- and He-dominated scenarios is through the fractionation of their outflow. Outflow observations mostly trace the upper atmosphere, yet the fractionation is controlled by the coupling in the lower atmosphere \citep{Xing2023}.

Numerous giant exoplanets, incl. HD~209458~b, present an outflow with a subsolar He/H ratio \citep[e.g.,][]{Lampon2020a,Lampom2021,Khodachenko2021,Shaikhislamov2021}. While these subsolar He/H outflow ratio were originally linked to a low atmospheric abundance of helium,  \citet{Xing2023} recently showed using a self-consistent multi-fluid hydrodynamic model that He-mass fractionation can explain the outflow of HD~209458~b. In other words, HD~209458~b does not need to be He-poor to explain the low He/H ratio of its outflow. In fact, the opposite can be true as it can preferentially lose hydrogen over billions of years.

\subsubsection{Atmospheric Chemistry}

A third avenue to considering leveraging to help disentangle the H$_2$- and He-dominated scenarios relates to the different chemical composition expected for each of them. One of the most notable differences/markers expected relates to CO and CO$_2$ being the main reservoir of carbon as CH$_4$ will have to compete with H$_2$O for the (relatively) scarce hydrogen \citep{Hu2015,Guzman2022}. The four molecules above being strong absorbers, their relative abundances may provide the most constraining means to disentangle between the two atmospheric scenarios--with the caveat that the interpretation of the abundances will be dependent (i.e., sensitive) to the atmospheric/chemistry models.

\subsection{Other Considerations for Robust Inferences}

\red{While no exoplanet spectra has been interpreted as sign of an unexpectedly-high metallicity (e.g., $O(10)\%$ water abundance) to date, a substantial fraction of interpretations invoke clouds \citep[see, e.g.,][and references therein]{Espinoza2025}. In addition, the abundances reported in some instances appear biased to low values due to priors. For example, while \citet{Bell2023} reports a $\log_{10}[\mathrm{MR_{H_2O}}] = -1.80^{+0.55}_{-0.94}$ ($1.6^{+4.0}_{-1.4}\%$) for WASP-80~b their PPD hits the upper bound of their priors well within 1$\sigma$ (see their Extended Data Fig. 3) meaning that without such hard edge the favored water abundance would likely be closer to $O(10)\%$. Jointly, these facts call for (re)analyses with a special care for the complex dependencies discussed here.}

\section{Conclusion}
\label{sec:conclusion}

\red{By freeing the helium-to-hydrogen ratio in atmospheric retrievals, we show that helium enrichment and high water content (or, more broadly, metallicity) can explain the data at hand---in addition to clouds.} This provides a new avenue for interpreting JWST spectra and upcoming composition/metallicity trends. It is also a reminder that as data quality increases, hidden assumptions in analysis frameworks may exacerbate biases in our inferences.

Moving forward, addressing hidden assumptions in atmospheric retrievals such as a fixed He/H$_2$ ratio, will be critical to reliably interpret the growing numbers of high-quality exoplanet spectra. As assumptions are lifted, complementary datasets will be essential to avoid drowning in subsequent degeneracies. 

We suggest here to leverage complementary insights from differences in (1) pressure-broadening behaviors, (2) outflow fractionation, and (3) atmospheric chemical makeup (with a focus on the main carrier of carbon) to disentangle between the He-rich and (H$_2$-dominated) volatile-rich cases--which we expect to require higher-fidelity opacity/outflow/chemistry models. These improvements will not only refine/consolidate metallicity measurements for hot Jupiters and mini-Neptunes, but will also help develop a more nuanced understanding of the atmospheric diversity and evolutionary pathways of exoplanets---incl., maybe, via the first detection of a He-dominated gas giant.

\section*{Acknowledgements}
The authors thank \red{the Reviewer}, C. Cherubim, \red{J. Fortney}, I. Gordon, R. Hargreaves, \red{Z. Lin}, S. Seager, \red{D. Thorngren}, L. Wiesenfeld, R. Wordworth, and Q. Xue for valuable inputs. The authors acknowledge the MIT SuperCloud and Lincoln Laboratory Supercomputing Center for providing HPC resources that have contributed to the present work. J.d.W. and P.N. acknowledge support from the European Research Council (ERC) Synergy Grant under the European Union’s Horizon 2020 research and innovation program (grant No. 101118581 — project REVEAL). 

\bibliography{bibliography}{}
\bibliographystyle{aasjournal}

\clearpage

\end{document}